\documentclass[useAMS,usenatbib]{mn2e}

\usepackage{graphicx}

\title[Frequencies and resonances around $L_4$]{Frequencies and resonances around $L_4$ in the elliptic restricted three-body problem}

\author[R. Rajnai, I. Nagy, B. \'Erdi]{R. Rajnai$^{1}$\thanks{E-mail:
rajnair@sze.hu}, I. Nagy$^{2}$, B. \'Erdi$^{3}$\\
$^{1}$Department of Mathematics and Computational Sciences, Sz\'echenyi University, Egyetem t\'er 1, H-9026 Gy\H{o}r, Hungary\\
$^{2}$Department of Science, National University of Public Service, Hung\'aria k\"or\'ut 9-11, H-1101 Budapest, Hungary\\ 
$^{3}$Department of Astronomy, E\"otv\"os University, P\'azm\'any P\'eter s\'et\'any 1/A, H-1117 Budapest, Hungary}

\begin{document}

\date{date}

\pagerange{\pageref{firstpage}--\pageref{lastpage}} \pubyear{2002}

\maketitle

\label{firstpage}

\begin{abstract}

The stability of the Lagrangian point $L_4$ is investigated in the elliptic restricted three-body problem by using Floquet's theory. Stable and unstable domains are determined in the parameter plane of the mass parameter and the eccentricity by computing the characteristic exponents. Frequencies of motion around $L_4$ have been determined both in the stable and unstable domains and fitting functions for the frequencies are derived depending on the mass parameter and the eccentricity. Resonances between the frequencies are studied in the whole parameter plane. It is shown that the 1:1 resonances are not restricted only to single curves but extend to the whole unstable domain. In the unstable domains longer escape times of the test particle from the neighbourhood of $L_4$ are related to certain resonances, but changing the parameters the same resonances may lead to faster escape.    

\end{abstract}


\section{Introduction}

The elliptic restricted three-body problem (ERTBP) is a thoroughly investigated problem of celestial mechanics which has received continuing attention for a long time, due to its theoretical interest and practical applicability in the dynamics of planetary systems. In the ERTBP a basic question is the stability of the Lagrangian triangular equilibrium point $L_4$. Since the ERTBP has two free parameters, the mass ratio $\mu$ and the orbital eccentricity $e$ of the primaries, and the linear variational equations of motion around $L_4$, determining stability or instability, have periodic coefficients, the problem is more difficult than in the circular restricted three-body problem, with the only parameter $\mu$ and constant coefficients of the variational equations. 

\par

Floquet's theory \citep{b9} of differential equations with periodic coefficients was used by \citet{b3} to determine the boundaries of the stable domain of $L_4$ in the $\mu,e$ plane, and by \citet{b1} to describe the structure of the unstable domain. \citet{b8} derived an algebraic equation from which the four frequencies of libration around $L_4$ in the ERTBP can be computed for small values of $e$. Transition curves, separating stable and unstable domains, were determined by analytical \citep{b11} and numerical \citep{b12} methods.

\par

Non-linear stability of $L_4$ and the extension of the stable region around $L_4$ in the configuration plane, depending on $\mu$ and $e$, were studied numerically by \citet{b7} and also by \citet{b13} who showed the shrinking of the stable region around $L_4$ at certain combinations of $\mu$ and $e$ corresponding to unstable resonances. Types of resonances between the frequencies of libration in the ERTBP were defined in \citet{b2} and their connection with the minima of the size of the stable region were studied in detail in \citet{b2,b4}. Applications to possible Trojan exoplanets were investigated in \citet{b20}. Recently, by using Hill's equation and the energy-rate method, \citet{b5} pointed out the dominant role of the long-period component of libration in forming the structure of the stability domain of $L_4$.

\par

In this paper we study the resonances between the frequencies of motion around $L_4$ in the ERTBP in a larger domain of the $\mu,e$ parameter plane than before \citep{b2,b4}, giving a more extended view on the relation between resonances and the structure of the stable and unstable domains of $L_4$.

\par

The paper is organized as follows. The equations of motion are given and the applied methods are described in Section \ref{SectionMethods}. Characteristic roots and characteristic exponents are discussed in Section \ref{SectionCharREx}. Comparison of the frequencies obtained by Floquet's theory and fast Fourier transformation is made in Section \ref{SectionFreqs}, where fitting functions for the frequencies are also determined. Resonances between the frequencies and their connection with the structure of the stable and unstable domains are studied in Section \ref{SectionRes}. Conclusions are drawn in Section \ref{Summary}. The Appendix gives the fitting functions and their coefficients for the frequencies.

\section[Method]{Model and method}

\label{SectionMethods}

\subsection{Equations of motion}

\label{EqsMot}

The ERTBP studies the motion of a point-like body with negligible mass (in the following a test particle), moving in the orbital plane and under the gravitational influence of two point-like massive bodies (the primaries), which revolve around their barycentre in elliptic orbits. The equations of motion of the test particle, in a barycentric coordinate system, rotating together with the primaries with the $x$-axis going through them,  are \citep{b19}

\begin{equation}
 x'' - 2y' = \alpha \frac{\partial \Omega}{\partial x}, 
           \quad
 y'' + 2x' = \alpha \frac{\partial \Omega}{\partial y}, 
\label{EqMo}
\end{equation} 

where $x$ and $y$ are the rectangular coordinates of the test particle, the prime means derivation according to the true anomaly $v$ of the primaries, serving as independent variable, 

\begin{displaymath}
\alpha = \frac{1}{1+e\cos v},
\end{displaymath}

and $e$ is the eccentricity of the relative orbit of the primaries.

The potential function $\Omega$ is

\begin{displaymath}
\Omega=\frac{1}{2}\left[(1-\mu)r_1^2+\mu r_2^2 \right]
+\frac{ 1-\mu}{r_1}+\frac{ \mu}{r_2},
\end{displaymath}

where $\mu$ is the mass parameter

\begin{displaymath}
\mu=\frac{m_2}{m_1+m_2},
\end{displaymath}

$m_1$ and $m_2$ being the masses of the primaries, and $r_1$, $r_2$ are the distances of the test particle from the primaries

\begin{displaymath}
r_1=\sqrt{(x-\mu)^2 + y^2}, \quad r_2=\sqrt{(x+1-\mu)^2 + y^2}.
\end{displaymath}

The coordinates and distances are dimensionless, the instantaneous distance between the primaries serving as distance unit (the equations of motion are written in the so-called rotating 'pulsating' coordinate system). The ERTBP depends on two parameters, the eccentricity $0 \leq e < 1$, and the mass parameter $0 < \mu \leq 0.5$.

\par

The equations of motion have five  equilibrium solutions, the Lagrangian points $L_i$. The linear stability of these points can be studied by using the first variational equations of motion 

\begin{equation}
  \left(
       \begin{array}{c}
             \xi'  \\
             \eta' \\
             \xi'' \\
             \eta''\\
        \end{array}
  \right)
=
  \left(
      \begin{array}{ccrc}
            0 & 0 & 1 & 0 \\
            0 & 0 & 0 & 1 \\
            \alpha \,\Omega_{xx}^{(i)} & \alpha \,\Omega_{xy}^{(i)} &  0 & 2 \\
            \alpha \,\Omega_{xy}^{(i)} & \alpha \,\Omega_{yy}^{(i)} & -2 & 0 \\
      \end{array}
  \right)
   \left(
      \begin{array}{c}
         \xi  \\
         \eta \\
         \xi' \\
         \eta'\\
       \end{array}
   \right),
\label{LinVarEq}
\end{equation}

where $\xi$, $\eta$, $\xi'$, $\eta'$ are infinitesimal displacements in the position and velocity coordinates of $L_i$, and the partial derivatives of $\Omega$ have to be computed at the points $L_i$. Specifically, $\Omega_{xx}^{(4)} = 3/4$, $\Omega_{xy}^{(4)} = \frac{3\sqrt{3}}{2}(\mu-1/2)$, $\Omega_{yy}^{(4)} = 9/4$ for the triangular Lagrangian point $L_4$. 

\par

For $e=0$, the system (\ref{EqMo}) gets simplified to the circular restricted three-body problem, for which Eqs (\ref{LinVarEq}) make up a linear system of differential equations with constant coefficients. For $e>0$, (\ref{LinVarEq}) is a linear system of differential equations with periodic coefficients (of period $2\pi$) that can be studied by using Floquet's theory \citep{b9}.

\subsection{Floquet's theory}

For convenience, here we repeat Floquet's theory, following the scenario described by \citet{b3} and \citet{b1}, based on \citet{b9}.

\par

By defining the new variables $x_1=\xi$, $x_2=\eta$, $x_3=\xi'$, $x_4=\eta'$, (\ref{LinVarEq}) can be written in the compact form 

\begin{equation}
\mathbf{x}' = \mathbf{A}(t) \mathbf{x} \ ,
\label{MatrixEq}
\end{equation} 

 where $\mathbf{x} = (x_1, x_2, x_3, x_4)^T$ (the upper index $ ^T$ denoting the transpose of a matrix), and $\mathbf{A}$ is the coefficient matrix in (\ref{LinVarEq}).

\par

Let $\mathbf{x}^1,\ldots, \mathbf{x}^4$ be $4$ linearly independent solutions of (\ref{MatrixEq}), then 

\begin{equation}
  \mathbf{X}(t) = \left(
     \begin{array}{cccc}
           \mathbf{x}^1 & \mathbf{x}^2 & \mathbf{x}^3 & \mathbf{x}^4 \\
     \end{array}
     \right)
\label{FundMatrix}
\end{equation}

is called a fundamental matrix, satisfying $\mathbf{X}'(t) = \mathbf{A}(t) \mathbf{X}(t)$.

\par

Floquet's theorem \citep{b9} states, that if $\mathbf{X}(t)$ is a fundamental matrix solution of the system (\ref{MatrixEq}) with periodic coefficients of period $T$, then so is $\mathbf{X}(t+T)$, and there exists a non-singular constant matrix $\mathbf{B}$, such that $\mathbf{X}(t+T) = \mathbf{X}(t)\mathbf{B}$ for all $t$. According to this, if $\mathbf{X}(t)$ is such a solution that at $t=t_0$, $\mathbf{X}(t_0) = \mathbf{I}$ where $\mathbf{I}$ is the identity matrix, then

\begin{equation}
\mathbf{X}(t_0+T) = \mathbf{B}.
\label{ConstBMatrix}
\end{equation}

\par

The eigenvalues $\lambda_j$ of $\mathbf{B}$ ($j=1,2,3,4$) are the characteristic roots of the system (\ref{MatrixEq}). They are an intrinsic property of the system, and  independent of the choice of the fundamental matrix. 

\par

The characteristic exponents $\nu_j$ are defined by 

\begin{equation}
\lambda_j = \exp{(\nu_j T)}, \quad j=1,2,3,4.
\label{CharRoot}
\end{equation}

Writing $\lambda_j$ in the form $\lambda_j = |\lambda_j|\cdot \exp(i \varphi_j)$, the characteristic exponents $\nu_j$ can be obtained from the characteristic roots $\lambda_j$ as 

\begin{equation}
\nu_j = \frac{\ln{|\lambda_j|}}{T} + i \left( \frac{\varphi_j + k2\pi}{T} \right), \quad k = 0, \pm 1, \pm 2, \ldots,
\label{CharExp}
\end{equation}

that is within a multiple of $2\pi i/T$.

\par

The consequence of Floquet's theorem is that there exist $4$ linearly independent solutions for (\ref{MatrixEq}) of the form 

\begin{equation}
\rmn{x}_j(t) = \exp{(\nu_j t)} p_j(t), \quad j=1,2,3,4,
\label{Solution}
\end{equation}

where $p_j(t)$ is a periodic function with period $T$.

\subsection{Applying Floquet's theory}

Applying Floquet's theory to study the stability of $L_4$, we integrated numerically Eqs (\ref{LinVarEq}) for a period of the primaries,  $T=2\pi$, with 4 different initial conditions at $t=t_0$:

\begin{displaymath}
\left(
\begin{array}{c}
x_1\\
x_2\\
x_3\\
x_4\\
\end{array}
\right)
=
\left(
\begin{array}{c}
1\\
0\\
0\\
0\\
\end{array}
\right)
,
\left(
\begin{array}{c}
0\\
1\\
0\\
0\\
\end{array}
\right)
,
\left(
\begin{array}{c}
0\\
0\\
1\\
0\\
\end{array}
\right)
,
\left(
\begin{array}{c}
0\\
0\\
0\\
1\\
\end{array}
\right).
\end{displaymath}

Thus $\mathbf{X}(t_0) = \mathbf{I}$. After one period, from the results of the numerical integration we could build up the matrix $\mathbf{B}$. Then we determined the characteristic roots $\lambda_j$ and exponents $\nu_j$ of the system by using GNU Octave's built in functions.

\par

We integrated Eqs (\ref{LinVarEq}) by changing the two parameters $\mu$ and $e$ in the regions $0 < \mu \le 0.5$, and $0 \le e < 1$, with stepsize $\Delta \mu= 0.0001$, $\Delta e= 0.005$, thus we determined the characteristic roots and exponents for the whole $\mu,e$ parameter plane.

\section[Characteristic roots and exponents]{Characteristic roots and exponents}

\label{SectionCharREx}

\subsection{Stability}

\label{SectionStability}

The types of the characteristic roots $\lambda_j$ determine the linear stability of $L_4$. It can be shown that the characteristic roots occur in reciprocal pairs (due to the fact that the equations of motion of the ERTBP can also be written in Hamiltonian form). Therefore, a pair of real roots  $\lambda_j$ and $1/\lambda_j$ always means instability, since either $|\lambda_j|>1$ or $1/|\lambda_j|>1$ and one solution (\ref{Solution}) becomes unbounded. Since the equation for the eigenvalues of the matrix $\mathbf{B}$ has real coefficients, complex characteristic roots appear in reciprocal and conjugate pairs. Thus in the case of complex characteristic roots, stability holds only if all roots have unit modulus $|\lambda_j| = 1$, that is they are on the unit circle (otherwise $|\lambda_j|>1$ or $1/|\lambda_j|>1$ would make the solution unbounded).

\begin{figure}

	\includegraphics[width=0.43\textwidth]{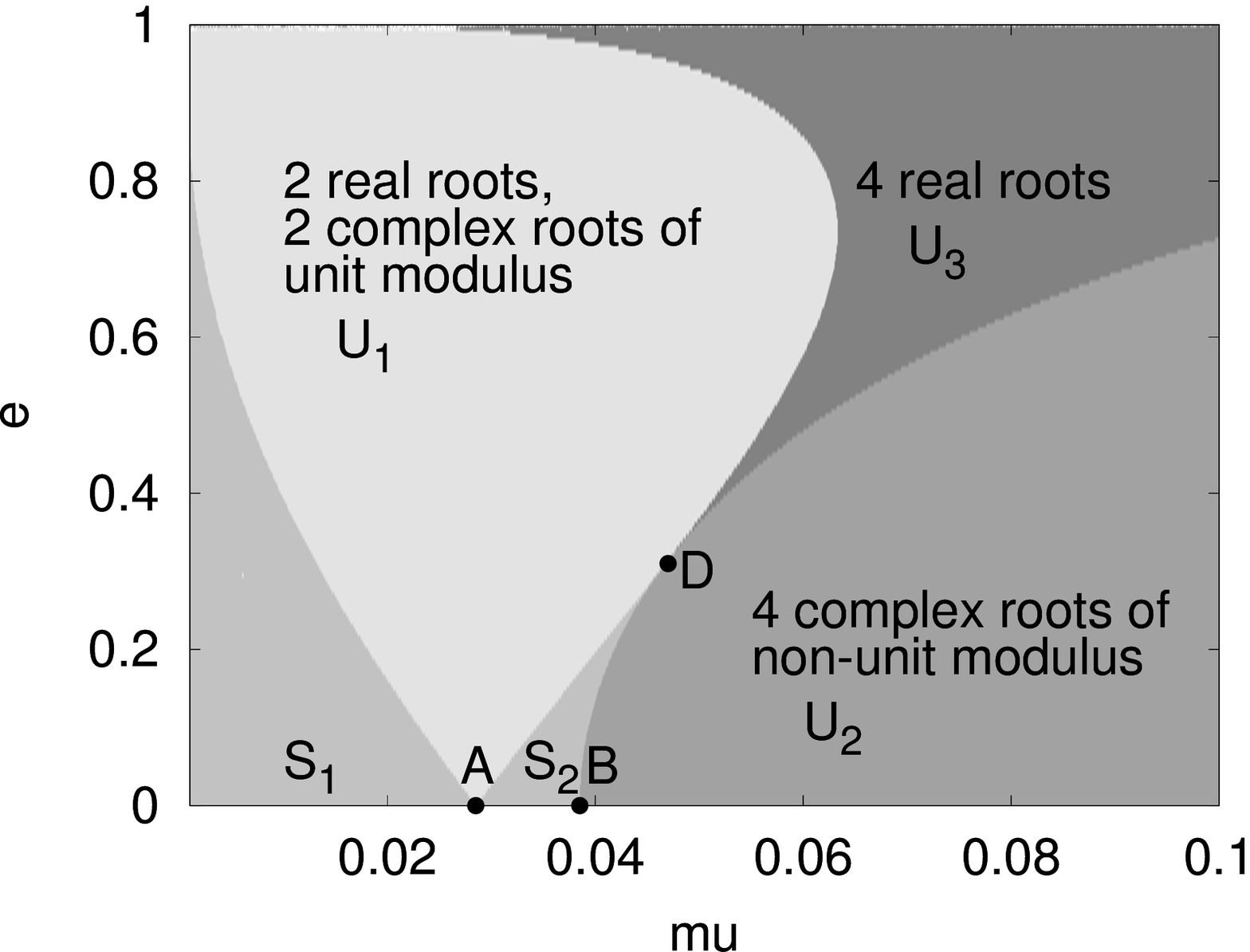}

	\includegraphics[width=0.43\textwidth]{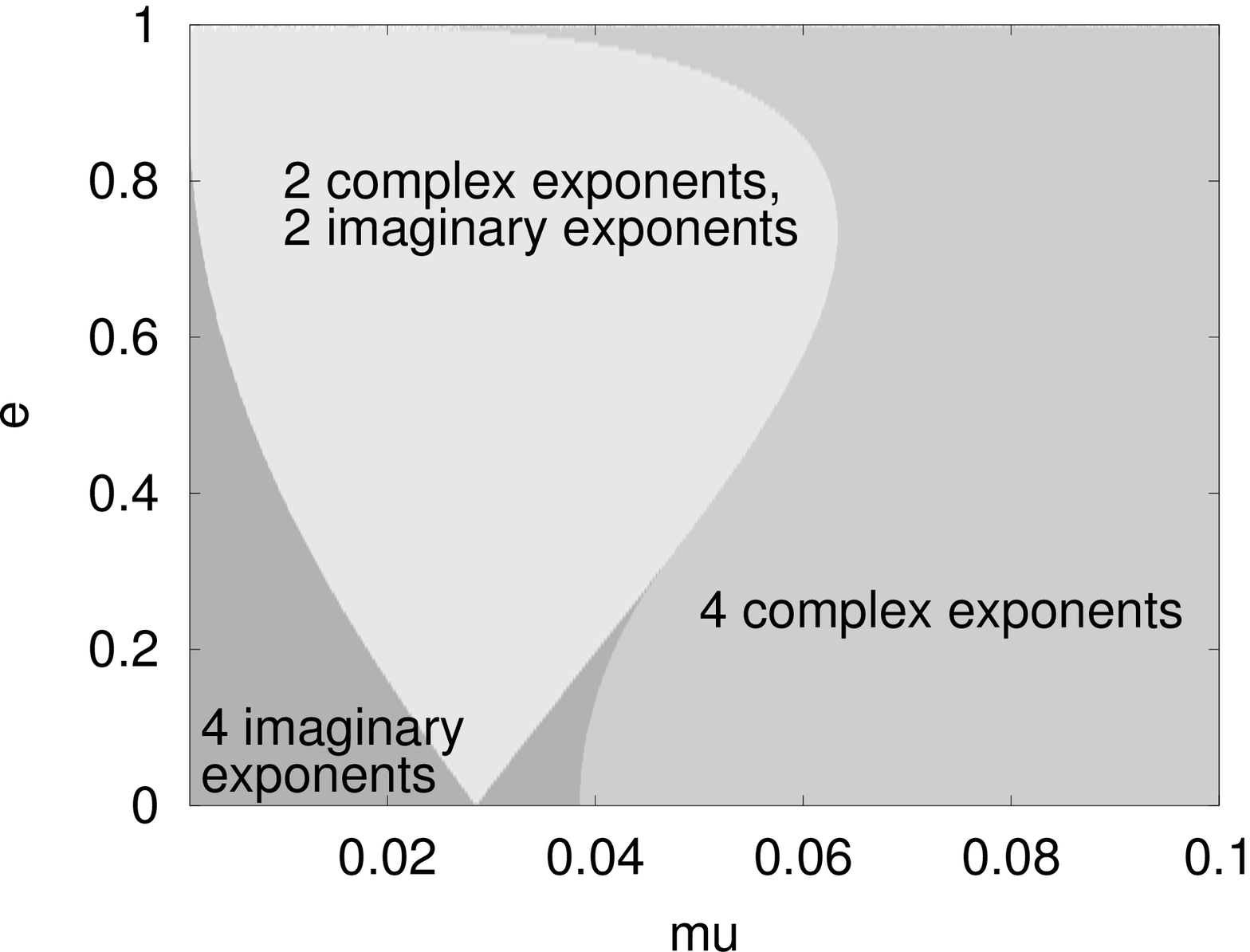}

\caption{The types of the characteristic roots (top panel) and characteristic  exponents (bottom panel) of the first variational equations, depending on $\mu$ and $e$. In the S1 and S2 regions (upper panel) there are 4 complex roots of unit modulus and $L_4$ is stable.}

\label{CharRoots}

\end{figure}

\par

We computed the characteristic roots depending on $\mu$ and $e$. According to their types, 4 regions can be distinguished in the $\mu,e$ plane as shown in the top panel of Fig. \ref{CharRoots} for $\mu<0.1$. This is in agreement with the results of \citet{b1}. $L_4$ is stable and stable periodic motion around it is only possible in the domain of 4 complex roots of unit modulus (S1 and S2 regions). The shape of the stability domain is well-known from previous investigations \citep{b3,b2}. Three critical points are associated with this region; two on the $e=0$ axis at $\mu=0.02859$ (point A in Fig. \ref{CharRoots}), dividing the stability domain for two parts, and at $\mu=0.03852$ (point B), limiting the stability domain on the $\mu$-axis, and a third at $e=0.3143$, $\mu=0.04698$ (point D) at the upper right peak of the stability domain.

\par

$L_4$ is unstable in three domains (marked by U1, U2, and U3 in the top panel of Fig. \ref{CharRoots}), with characteristic roots of different properties. There are 2 real roots and 2 complex roots of unit modulus in the U1 domain, 4 complex roots of non-unit modulus in U2, and 4 real roots in U3. The U2 and U3 domains extend up to $\mu=0.5$, the border between them approaching to $e=1$ (not shown in Fig. \ref{CharRoots}).

\par

The bottom panel of Fig. \ref{CharRoots} shows the types of the characteristic exponents $\nu_j$ in the $\mu,e$ plane. In the stable domain, there are 4 purely imaginary exponents corresponding to the 4 complex characteristic roots of unit modulus in the S1, S2 domains. To the characteristic roots in the U1 domain, there correspond 2 complex and 2 imaginary exponents in the bottom panel of Fig. \ref{CharRoots}. The boundary between the U2 and U3 domains on the top panel vanishes in the bottom panel, since the 4 real characteristic roots are all negative and 4 complex exponents correspond to them, as well as to the roots in the U2 domain. However, the complex exponents corresponding to the roots in the U3 domain have equal imaginary parts, while the exponents corresponding to the roots in the U2 domain have two pairs of equal imaginary parts. 

\par

Moving away from the boundary of the stable domain (S1, S2), with increasing the eccentricity or the mass parameter, the characteristic roots slowly drift away from the unit circle, and small real parts appear in the characteristic exponents, causing the solution (\ref{Solution}) to become unstable. The dissolution of the system (the escape of the test particle from $L_4$) is slower near the boundary of the stable domain, and speeds up with the increase of the positive real parts of the characteristic exponents. This is in good agreement with the escape times from $L_4$, computed in \citet{b4}.

\par

\begin{figure}

	\includegraphics[width=0.34\textwidth, angle=270]{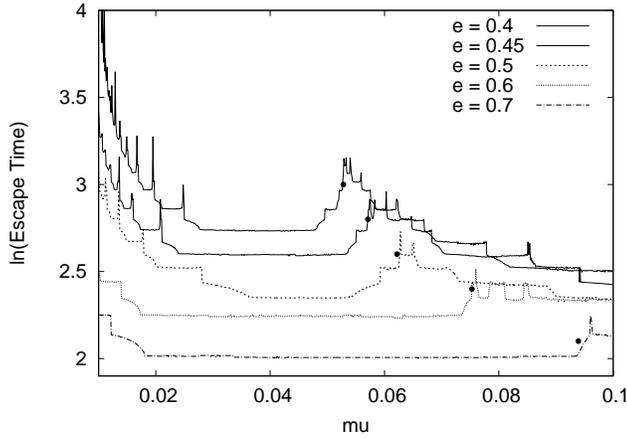}

\caption{The escape time of the test particle from $L_4$ (on a logarithmic scale) depending on $\mu$, for several values of $e \geq 0.4$. (We note that for $e > 0.3143$ $L_4$ is unstable for most values of $\mu$, see Fig. \ref{CharRoots}.)  Black dots mark the boundary points between the U2 and U3 domains of Fig. \ref{CharRoots} for the same values of $e$.}

\label{CharRootsVsLifetime}

\end{figure}

In Fig. 2 of \citet{b4}, the unstable region is divided into two parts by a dim boundary, along which the lifetime of the system (until the test particle remains in the vicinity of $L_4$) slightly increases. The authors suspected that the boundary between the U2 and U3 domains is responsible for this. Here, in Fig. \ref{CharRootsVsLifetime}, we show the escape time of the test particle (on a logarithmic scale) as a function of $\mu$, for several values of $e$, determined by \citet{b10}. For the same values of $e$, the points of the boundary between the U2 and U3 domains of  Fig. \ref{CharRoots} are also marked (black dots). It can be seen that these boundary points fit well to the places of the slight local increase of the escape time. As we shall see in Section \ref{SectionRes}, this boundary corresponds to several 1:1 resonances, making resonances accountable for the small growth in the escape time.

\subsection{Characteristic exponents}

\label{SectionChExp}

The real parts of the characteristic exponents $\nu_j$ in (\ref{CharExp}) are responsible for the exponential escape of the test particle from $L_4$, while the imaginary parts  in (\ref{CharExp})  result in periodic motion of infinitesimal amplitude around $L_4$ with frequencies

\begin{equation}
n_j= \frac{\displaystyle \varphi_j + k2\pi}{\displaystyle T}.
\label{ChFreq}
\end{equation}

Considering that $\nu_j$ occur in complex conjugate pairs (similarly to the complex characteristic roots $\lambda_j$), and that in the ERTBP there are 4 frequencies of libration around $L_4$ \citep{b8,b2}, we specified $k=0,1$ in (\ref{ChFreq}). Thus with $T=2\pi$, the 4 frequencies in the ERTBP are $n_s$, $n_l$, $1-n_s$, and $1-n_l$, where $n_s$ and $n_l$ are the frequencies of the short and long period libration around $L_4$, while $1-n_s$ and $1-n_l$ are due to the elliptic motion of the Lagrangian point $L_4$ itself.

\begin{figure}

	\includegraphics[width=0.45\textwidth]{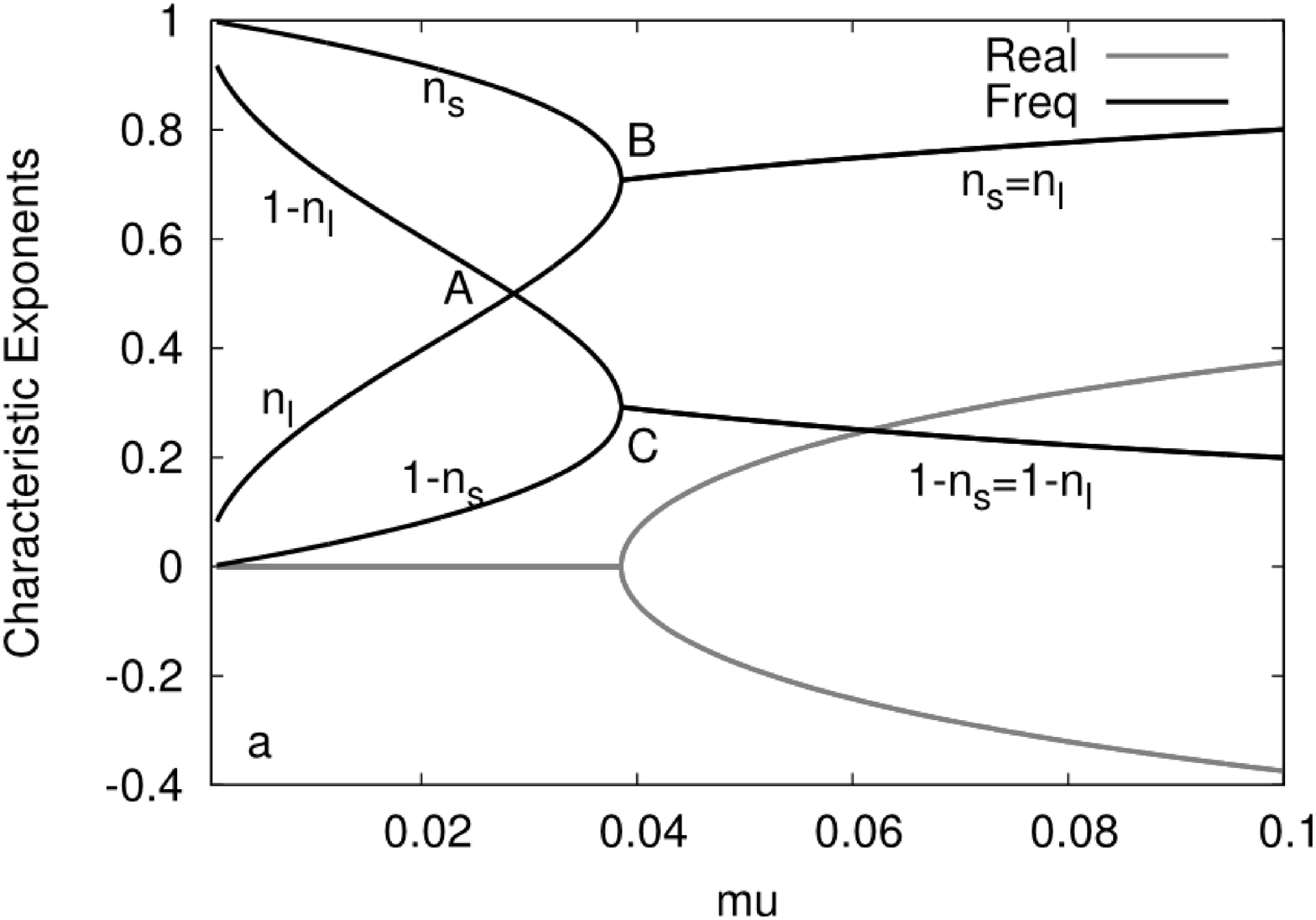}

	\includegraphics[width=0.45\textwidth]{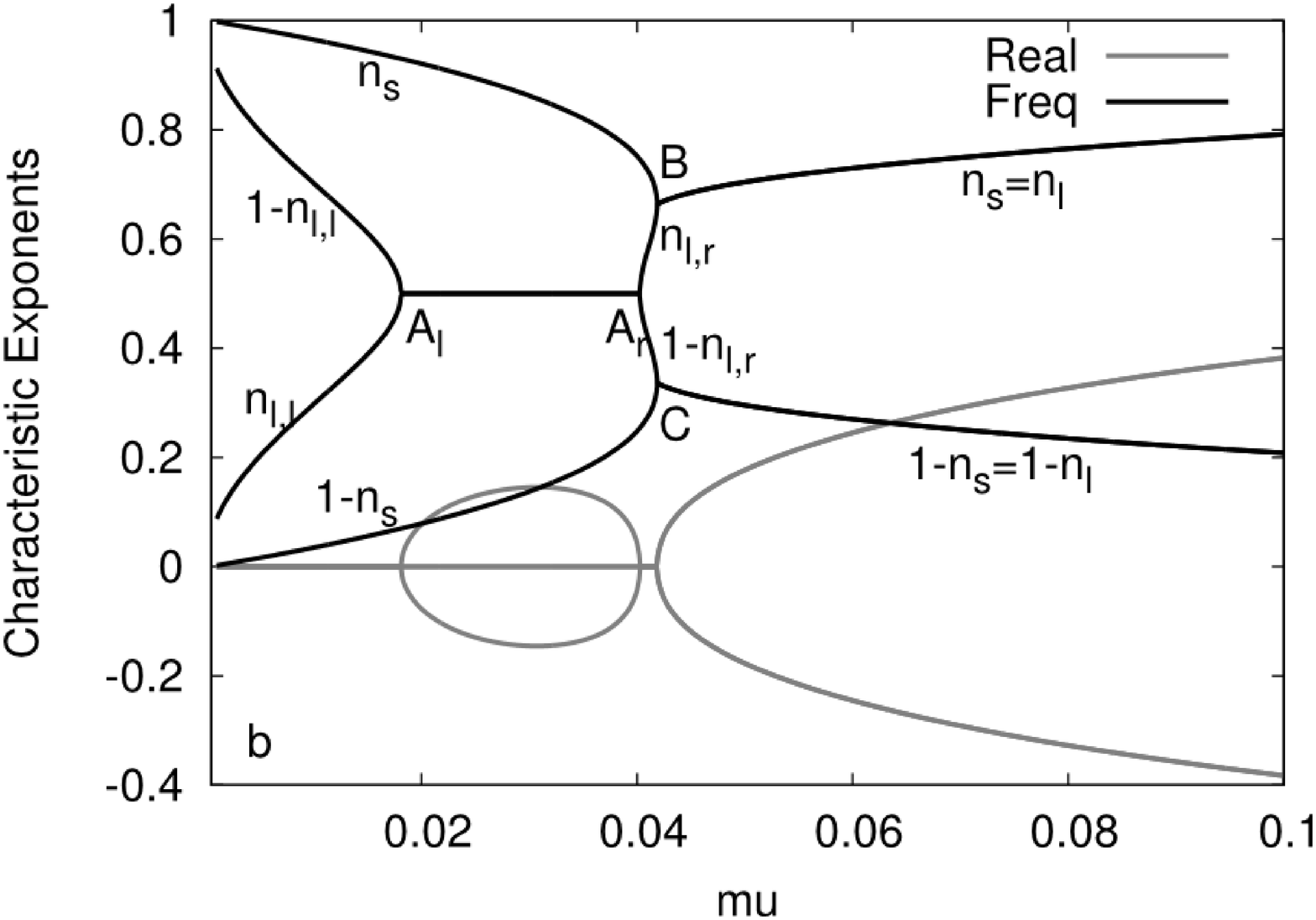}

	\includegraphics[width=0.45\textwidth]{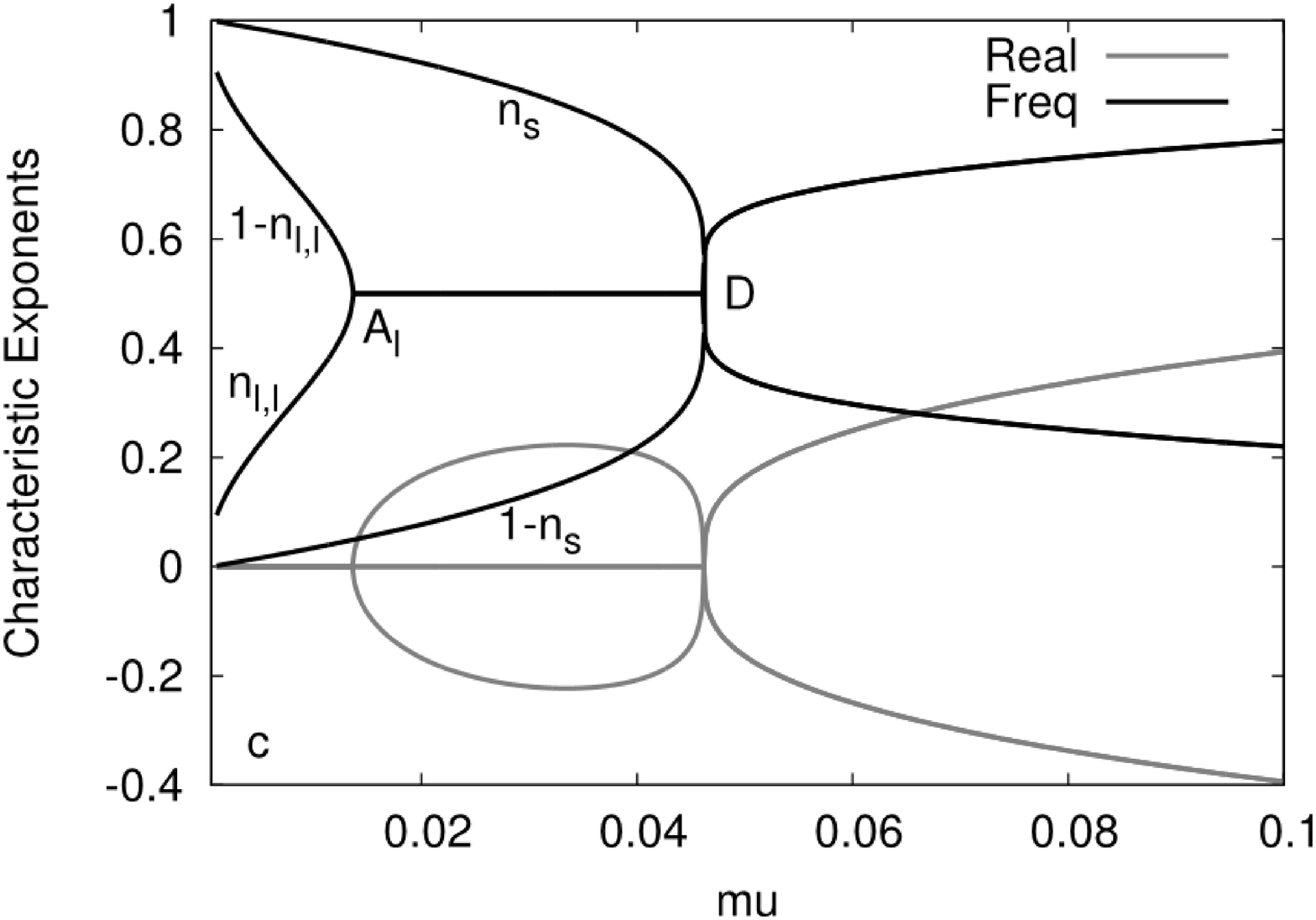}

  \includegraphics[width=0.45\textwidth]{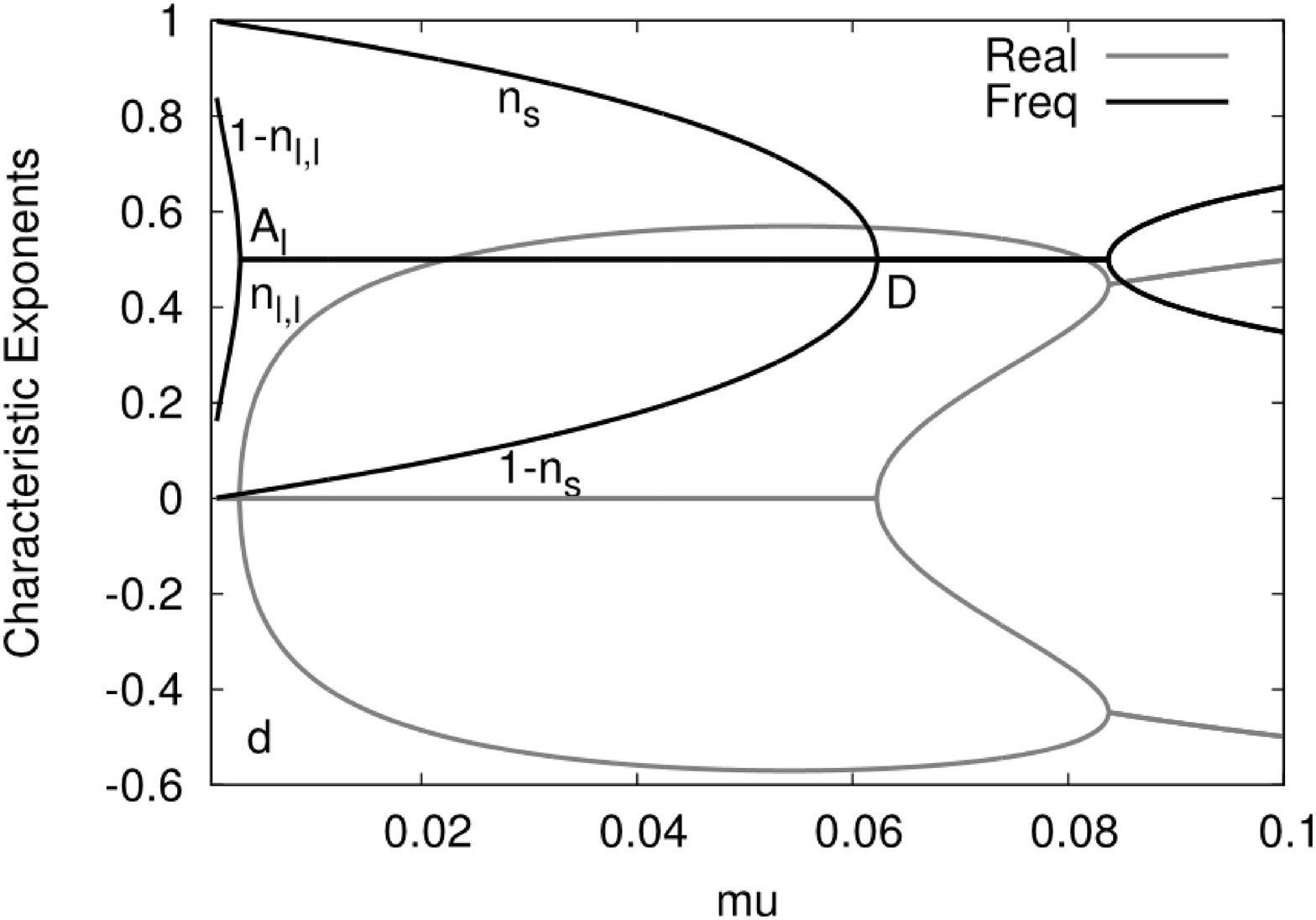}

\caption{The real parts and the frequencies corresponding to the imaginary parts of the characteristic exponents, depending on $\mu$, for several values of $e$: panel (a) $e=0$; (b) $e=0.2$; (c) $e=0.3$; (d) $e=0.65$.}

\label{ImagExp}

\end{figure}

\par

Fig. \ref{ImagExp} shows the real parts and the frequencies corresponding to the imaginary parts of the characteristic exponents, depending on $\mu$, for several values of $e$. 

The panel (a) displays the real parts (gray curves) and the fequencies  (black curves) for the limit case $e = 0$. For $\mu < 0.03852$, there are 4 frequencies, $n_s$, $n_l$, $1-n_s$, $1-n_l$. The point A marks the place, where the frequencies $n_l$ and $1-n_l$ become equal ($n_l=0.5$). This occurs at $\mu=0.02859$, at the critical point A of the stability domain in Fig. \ref{CharRoots}. The point A is the starting point of the A11 type resonance for $e>0$, as it will be seen in Section \ref{SectionRes}. Similarly, the points B and C are the starting points of the B11 and C11 resonances, where $n_s=n_l$  and $1-n_l=1-n_s$, respectively. These points occur at $\mu=0.03852$, corresponding to the point B of the stability domain in Fig. \ref{CharRoots}. 

\par

The difference of the present work from earlier ones \citep{b3,b1,b2} is, that  from the characteristic exponents we computed the frequencies in such domains of the $\mu,e$ plane, where they were not considered before. Thus  increasing the mass parameter over $\mu = 0.03852$, in the unstable domain the frequencies remain equal in pairs, $n_s = n_l$ and $1-n_l = 1-n_s$ (Fig. \ref{ImagExp}a), and instead of the so-far zero real parts in the stable domain ($\mu < 0.03852$), 2 non zero real parts appear, causing the test particle to escape from $L_4$.

\par

Increasing the eccentricity, the point A of Fig. \ref{ImagExp}a splits into two, as can be seen in Fig. \ref{ImagExp}b for $e=0.2$ (A$_l$ on the left, A$_r$ on the right). For values of $\mu$ between A$_l$ and A$_r$, there are two complex and two imaginary characteristic exponents (see the bottom panel of Fig. \ref{CharRoots}). The frequencies corresponding to the  imaginary parts of the 2 complex characteristic exponents are equal, $n_l=1-n_l=0.5$, and can be seen between A$_l$ and A$_r$  (Fig. \ref{ImagExp}b). 

To the 2 imaginary characteristic exponents there correspond the $n_s$ and $1-n_s$ frequencies. For values of $\mu$ in the stability region, on the left from A$_l$ and on the right from A$_r$ there are 4  frequencies, corresponding to the 4 imaginary characteristic exponents. In the unstable domain ($\mu > 0.03852$), $n_s = n_l$ and $1-n_l = 1-n_s$, as in Fig. \ref{ImagExp}a. The properties of the real parts are similar to those of Fig. \ref{ImagExp}a, and two more real parts appear for values of $\mu$ between A$_l$ and A$_r$, corresponding to the two complex characteristic exponents in this domain (see the bottom panel of Fig. \ref{CharRoots}).  

\par

Increasing further the eccentricity, the points A$_l$ and A$_r$ move in opposite derection, B and C move toward each other, and finally for $e=0.3143$ and $\mu=0.04698$ the three points  A$_r$, B, and C are united in one point, marked by the letter D in Fig. \ref{ImagExp}c, and corresponding to the upper right peak D of the stability domain in Fig. \ref{CharRoots}, where all 4 frequencies become equal. Afterwards, the properties of the frequencies, and that of the real parts in the whole domain of $\mu$ are similar to as in Fig. \ref{ImagExp}b. 

\par

In Fig. \ref{ImagExp}d one can see the frequencies for $e=0.65$. Here D corresponds to that point, where in the upper panel of Fig. \ref{CharRoots} the $e=0.65$ line would cross the border between the U1 and U3 domains. All frequencies are equal from the point D until another point which would correspond to the intersection of the $e=0.65$ line and the border of the U2 and U3 domains. From this point on, two pairs of equal frequencies exist. For larger values of $\mu$ ($0.1<\mu<0.5$), the properties of the frequencies and the real parts are similar to as in Fig. \ref{ImagExp}d. 

\par

We note that for a given eccentricity, the real parts of the characteristic exponents change very steeply near the border of the stability regions S1 and S2 (see Figs \ref{ImagExp}b, c, d for values of $\mu$ corresponding to the point A$_l$, for example), and the abrupt changes in the positive real parts result in shorter escape times at the border of the stable region.

\section[Frequencies]{Frequencies} 

\label{SectionFreqs}

\subsection{A comparison of the frequencies}

\label{SectionFFT}

In Section \ref{SectionChExp} we determined the frequencies of motion around $L_4$ by using Floquet's theory. For comparison, we also computed the frequencies by the method of fast Fourier transform (FFT). 

For this we integrated the equations of motion of the ERTBP  numerically, over 1250 periods of the primaries (this proved to be a long enough time interval), by changing $e$ and $\mu$ on the same grid as in Section 2.3. The test particle was given a $10^{-6}$ displacement from $L_4$ in the $x$ direction as initial condition (and zero initial velocity in the rotating coordinate system). 

\par

\begin{figure}

	\includegraphics[width=0.42\textwidth, angle=270]{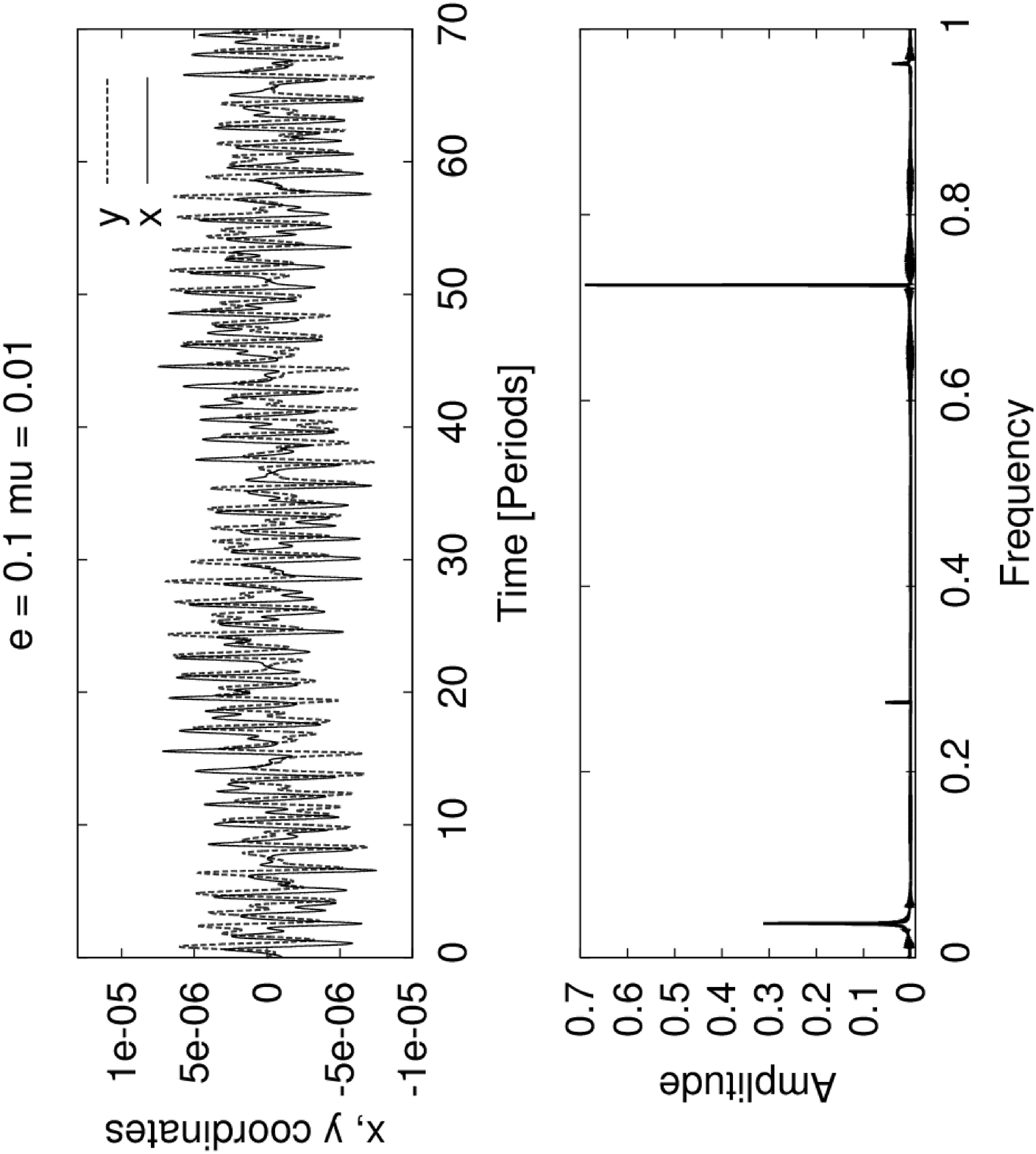}

\caption{Top panel: Variation of the $x$ and $y$ coordinates of the test particle relative to $L_4$ with time for $e = 0.1$, $\mu = 0.01$, and initial conditions described in the text. Bottom panel: Fourier spectrum of the motion shown in the top panel.}

\label{FFTSpectrum}

\end{figure}

The top panel of Fig. \ref{FFTSpectrum} shows the $x$ and $y$ coordinates of the test particle relative to $L_4$ for $e = 0.1$, and $\mu = 0.01$. Periodicities are well recognizable in both coordinates, the motion of the test particle is stable for this value of $e$ and $\mu$. Applying FFT on the time series of the coordinates, Fourier spectra can be determined on which 4 peaks, corresponding to the 4 frequencies of motion around $L_4$, can be identified (bottom panel of Fig. \ref{FFTSpectrum}).

\begin{figure}

	\includegraphics[width=0.34\textwidth, angle=270]{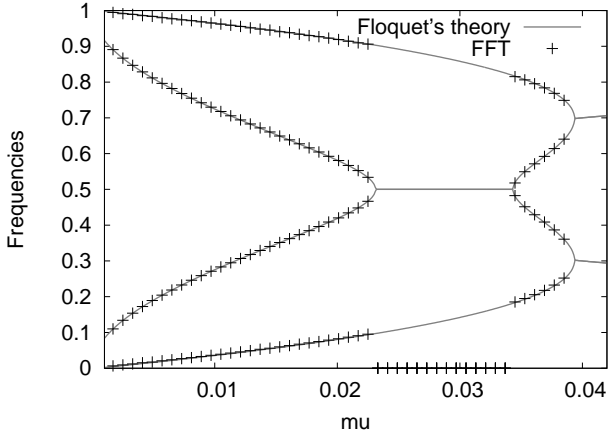}

\caption{Comparison of the frequencies determined by using FFT (marked by cross), and  Floquet's theory (solid lines) for $e = 0.1$. Between $0.021<\mu<0.033$ the frequencies refer to the unstable region U1.}

\label{FFT}

\end{figure}

\par

FFT performed well in the stable domain (S1 and S2), but in the unstable region (U1) the lifetime of the system was too short to get usable spectra. The decay of the system was due to escape or collision of the test particle with either of the primaries. In the former case exponential runaway dominated the motion, so the spectrum also. These effects are responsible for the missing FFT data in Fig. \ref{FFT}, where comparison of the two methods, FFT and Floquet's theory can be seen for $e=0.1$.  For this value of $e$, $L_4$ is unstable between $0.021<\mu<0.033$ (see Fig. \ref{CharRoots}), and there are 2 imaginary and 2 complex characteristic exponents. To the imaginary exponents there correspond periodic motions, however their frequencies could not be determined by FFT for the aforementioned reasons. According to Fig. \ref{FFT}, the agreement between the results of the two methods are good, and where FFT could not work, the application of Floquet's theory could provide the frequencies.

\subsection{Fitting the frequencies}

\label{SectionFit}

In Section \ref{SectionChExp} we determined the frequencies by using Floquet's theory. Fig. \ref{ImagExp} shows these frequencies for several values of $e$ as the function of $\mu$. In the limit case $e=0$, the $n_s$, $n_l$ frequencies in the stable domain $0<\mu<0.03852$ can also be computed from the well-known equations \citep{b4} 

\begin{eqnarray}
 & & n_s = \sqrt{| -0.5 - 0.5\sqrt{27(\mu-0.5)^2 -5.75} | }, \nonumber \\ [-5pt]
 & &                                                                   \\ [-5pt]
 & & n_l = \sqrt{| -0.5 + 0.5\sqrt{27(\mu-0.5)^2 -5.75} | }. \nonumber 
\label{LibFreq}
\end{eqnarray}

\par

Comparing the top and middle panels of Fig. \ref{ImagExp}, it can be seen that the frequencies are changing with the increase of the  eccentricity, however, the character of the curves remains. This gives the idea of searching for fitting functions of the frequency curves in the form of  Eq. (10). Due to the symmetry of the curves, it is enough to fit those parts of the curves which correspond to the $n_s$ and $n_l$ frequencies. From these the fit for the $1-n_s$, $1-n_l$ frequencies is immediately obtained. 

\par

The $n_s$ curve remains continuous as the eccentricity is increasing, but the $n_l$ curve has a breakpoint at $\mu=0.02859$, marked by a letter A in Fig. \ref{ImagExp}a. Increasing $e$ above $0$, the point A splits to two points (A$_l$ and A$_r$ in Fig. \ref{ImagExp}b). Thus the $n_l$ curve also splits to a left $n_{l,l}$ and a right $n_{l,r}$ side. 

(Between A$_l$ and A$_r$, $n_l=1-n_l=0.5$ as we have seen in Section \ref{SectionChExp}.) Therefore, we fitted the $n_{l,l}$ and  $n_{l,r}$ curves separately. Denoting the $\mu$ coordinates of the A$_l$ and A$_r$ points by $\mu_l$ and $\mu_r$, respectively, we fitted the $n_{l,l}$ curve in the interval $0<\mu<\mu_l$, and the $n_{l,r}$ curve in $\mu_r<\mu<0.04698$. For a given value of $e$, these are the two domains in $\mu$ where $L_4$ is stable. In the Appendix, we give the equations from which $\mu_l$ and $\mu_r$ can be computed for a given value of $e$.

\par

We assumed the fitting functions for the $n_s$, $n_{l,l}$, and $n_{l,r}$ curves in the form 

\begin{equation}
f(\mu) = a_0\sqrt{|a_1 +a_2\sqrt{a_3(\mu+a_4)^2 +a_5} | } +a_6,
\label{FitFunc}
\end{equation}

by allowing the dependence of the fitting parameters $a_i$ on the eccentricity,  $a_i = a_i(e)$. However, we got the best fits by keeping  $a_3 = 27.0$, $a_5 = -5.75$ for each frequency; $a_0 = 1.0$, $a_6 = 0.0$ for $n_s$ and $n_{l,l}$; and $a_0 = 0.5$ for $n_{l,r}$, independent of $e$. The remaining parameters showed dependence on $e$. These parameters were determined in the following way.

\par

\begin{figure}

	\includegraphics[width=0.43\textwidth,angle=270]{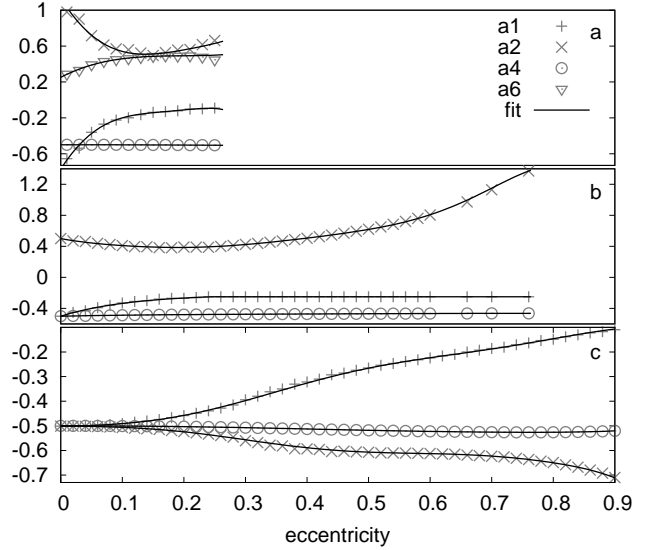}

\caption{The $a_i$ parameters depending on the eccentricity for the frequencies $n_{l,r}$ (a); $n_{l,l}$ (b); and $n_{s}$ (c) panel. Grey crosses, x-es, circles, and triangles indicate the determined parameter values for the given eccentricity, while the black solid lines mark the fitted polynomial functions.}

\label{Parameters}

\end{figure}

We computed the $n_s$, $n_{l,l}$, $n_{l,r}$ frequency curves for different values of $e$ by using Floquet's theory, and fitted them by the functions (\ref{FitFunc}). Fig. \ref{Parameters} shows the computed $a_i$ coefficients depending on $e$ (except the fixed ones, which do not depend on $e$). Fig. \ref{Parameters}a refers to the $n_{l,r}$ curves in the domain

$0<e<0.3143$, $\mu_r<\mu<0.04698$. The upper limits correspond to the upper right peak of the stability region (point D in Fig. \ref{CharRoots}). We note that near this peak the number of the determined frequencies were too low for an efficient fit. Fig. \ref{Parameters}b shows the coefficients for the $n_{l,l}$ curves in the domain $0<e<0.75$, $0<\mu<\mu_l$. (Due to the fast changes in the $n_l$ frequency for large values of $e$ we could determine the fitting coefficints only up to $e \sim 0.75$.) In  Fig. \ref{Parameters}c, the coefficients for the $n_s$ curves can be seen for $0<e<0.9$, $0<\mu<0.0633$. 

\par

Next we fitted the computed $a_i$ coefficients with polynomial functions of the eccentricity

\begin{displaymath}
a_i(e) = \sum_j a_{i,j} \cdot e^j,
\end{displaymath} 

except $a_1$ for the $n_{l,l}$ curve, where we used an exponential fitting function (based on several trials searching for the best representation). In Fig. \ref{Parameters} the black solid lines mark the fitted polynomial functions, whose $a_{i,j}$ coefficients are given in the Appendix.

\par

Fig. \ref{FitFreq} shows the frequencies computed by using Floquet's theory for several eccentricities (marked by different symbols), and obtained from the fitting functions (\ref{FitFunc}) with the determined coefficients (black solid lines). The accuracy of the fit of the $n _l$ curves decreases at the border  (at the A$_l$, A$_r$ points, see Fig. \ref{ImagExp}b) and near the upper right peak point D of the stability domain, due to the rapid changing of the frequency and the low number of the fitted data. However, farther from these places, the fit agrees well with the computed frequencies.

\begin{figure}

	\includegraphics[width=0.45\textwidth]{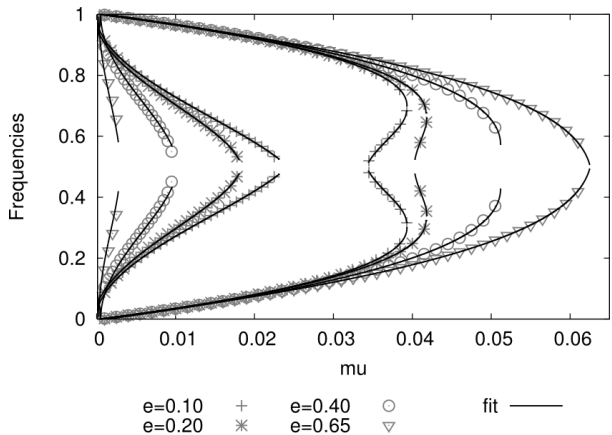}

\caption{Grey symbols mark the the frequencies computed by using Floquet's theory, and solid black lines stand for frequencies obtained from equation (\ref{FitFunc}), after fitting the parameters. The $n_l=1-n_l=0.5$ frequencies are not shown for clarity of the figure. }

\label{FitFreq}

\end{figure}

\section[Resonances]{Resonances}

\label{SectionRes}

\citet{b2} determined the size distribution of the stable regions of motions around $L_4$ in the $\mu,e$ plane and found that there are minimum zones, whose places can be related to resonances between the 4 frequencies of libration in the ERTBP. The possible resonances are defined in \citet{b2} and listed in Table \ref{Resonances}.

\par

\begin{figure}

	\includegraphics[width=0.47\textwidth]{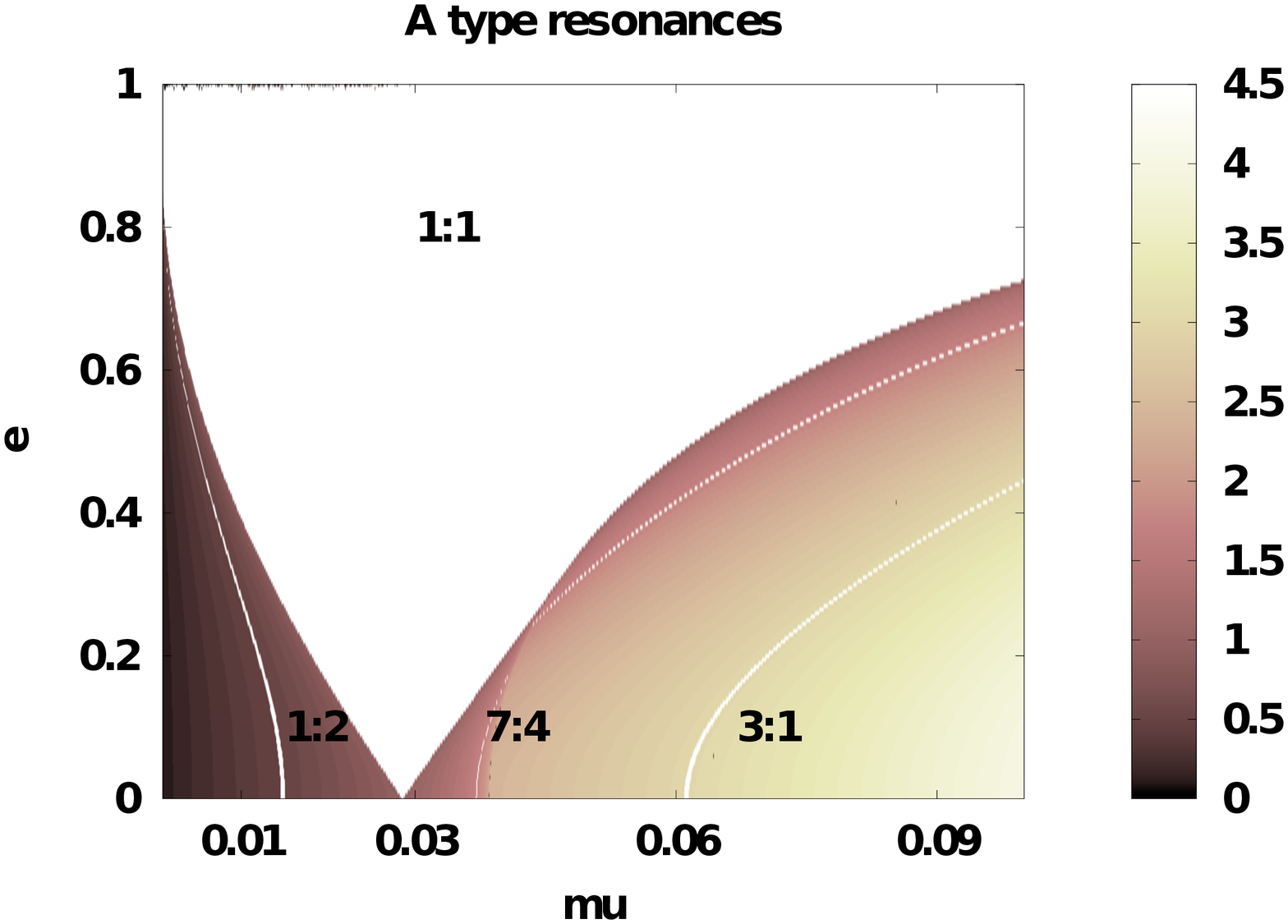}

	\includegraphics[width=0.47\textwidth]{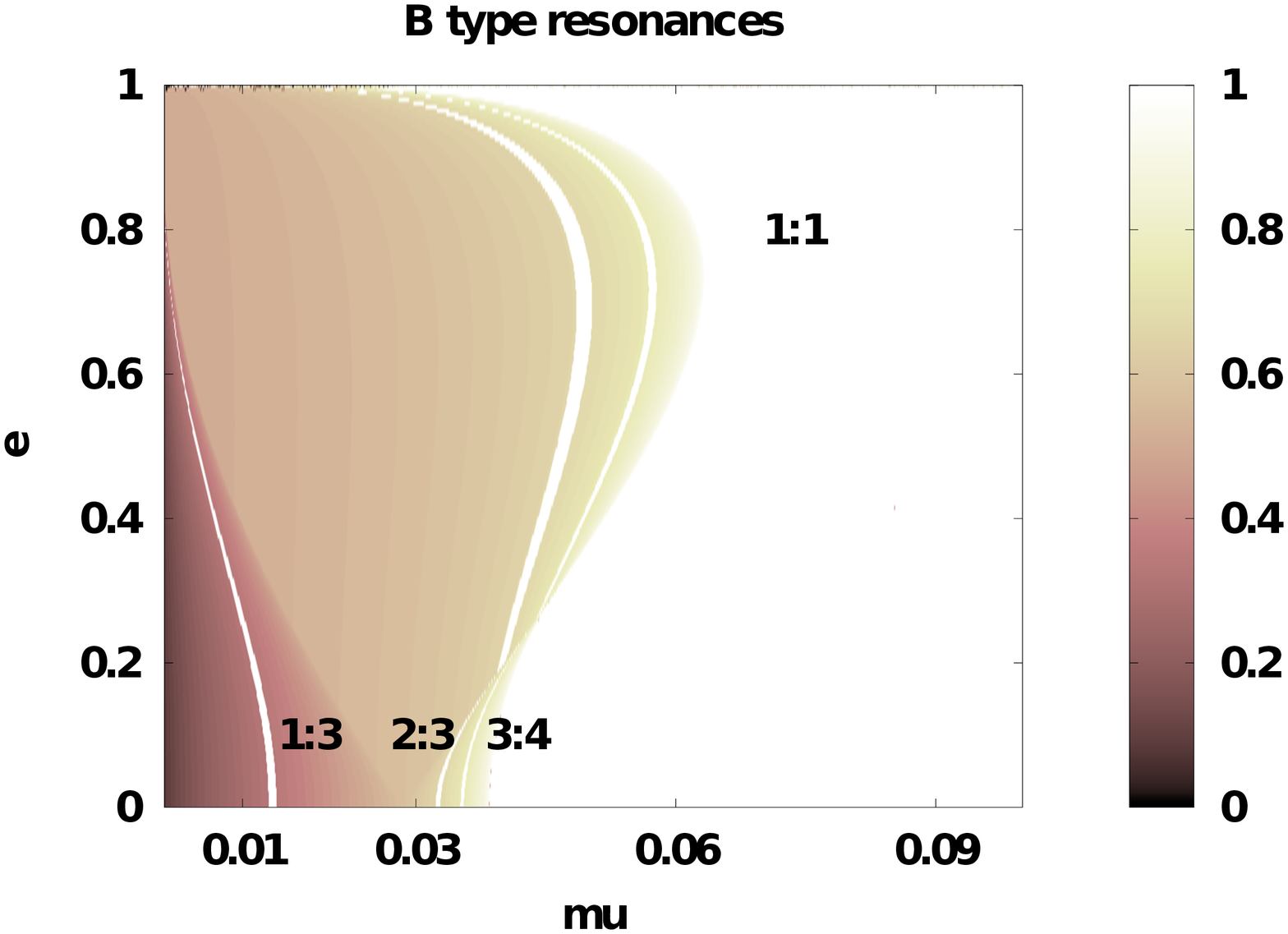}

	\includegraphics[width=0.47\textwidth]{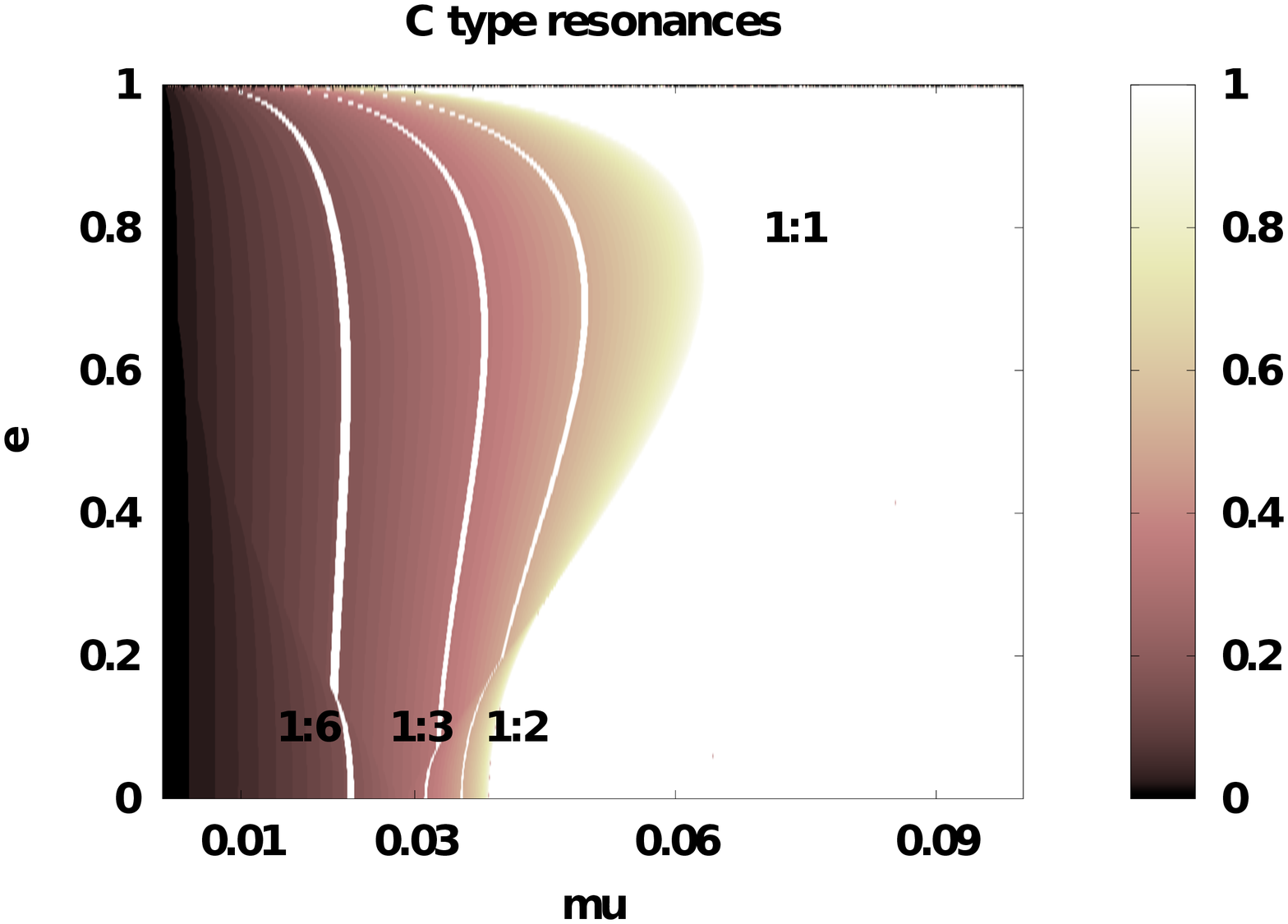}

\caption{Maps for the A, B, and C types of resonances. The colorbar indicates the continuous change of the reciprocal of the frequency ratios on the $\mu,e$ plane. The white curves indicate a few  resonances between the frequencies. The extended white domain in each panel stays for 1:1 resonances, where the frequencies are equal.}

\label{ABC}

\end{figure}

\begin{figure}

	\includegraphics[width=0.47\textwidth]{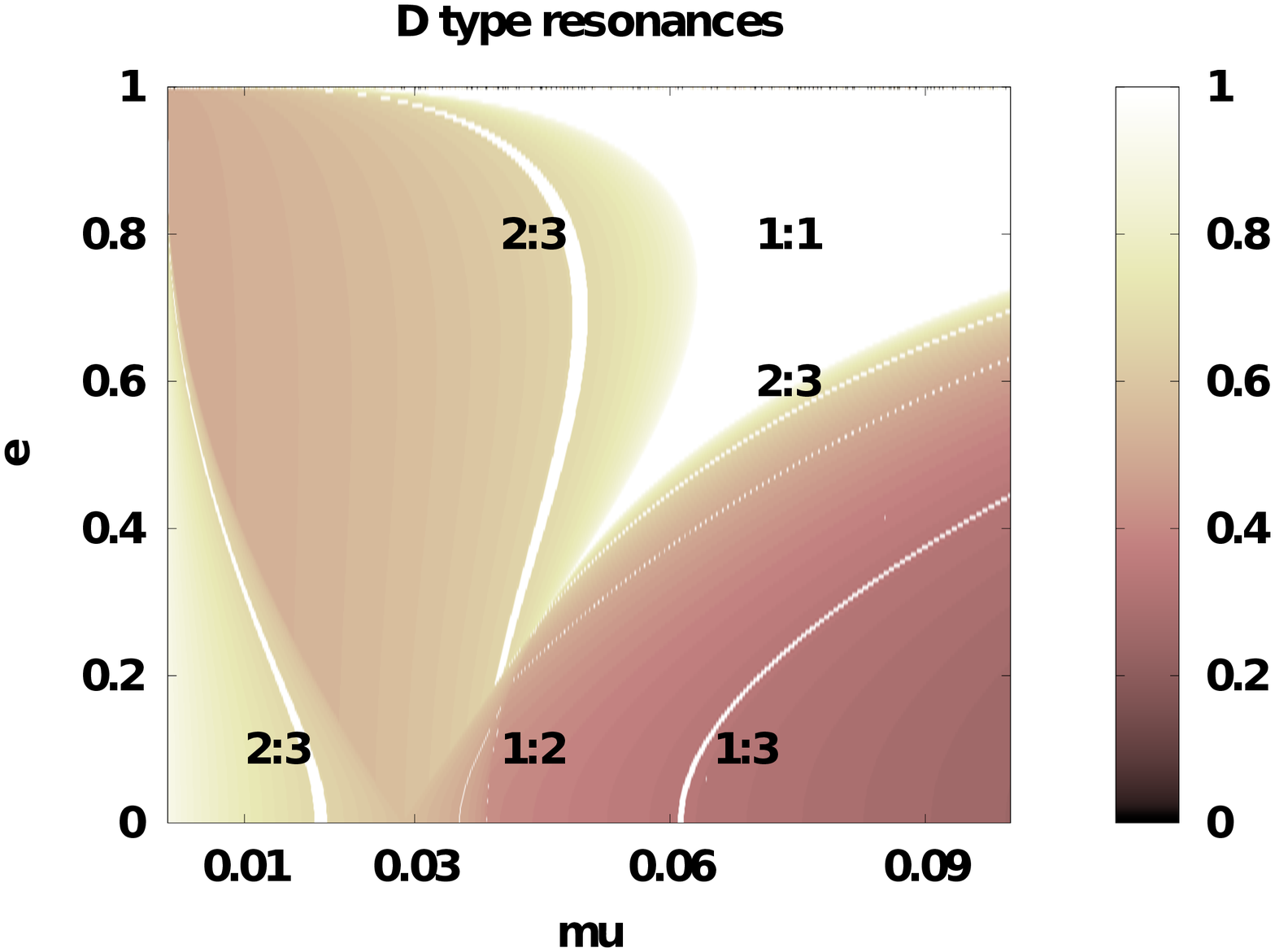}

	\includegraphics[width=0.47\textwidth]{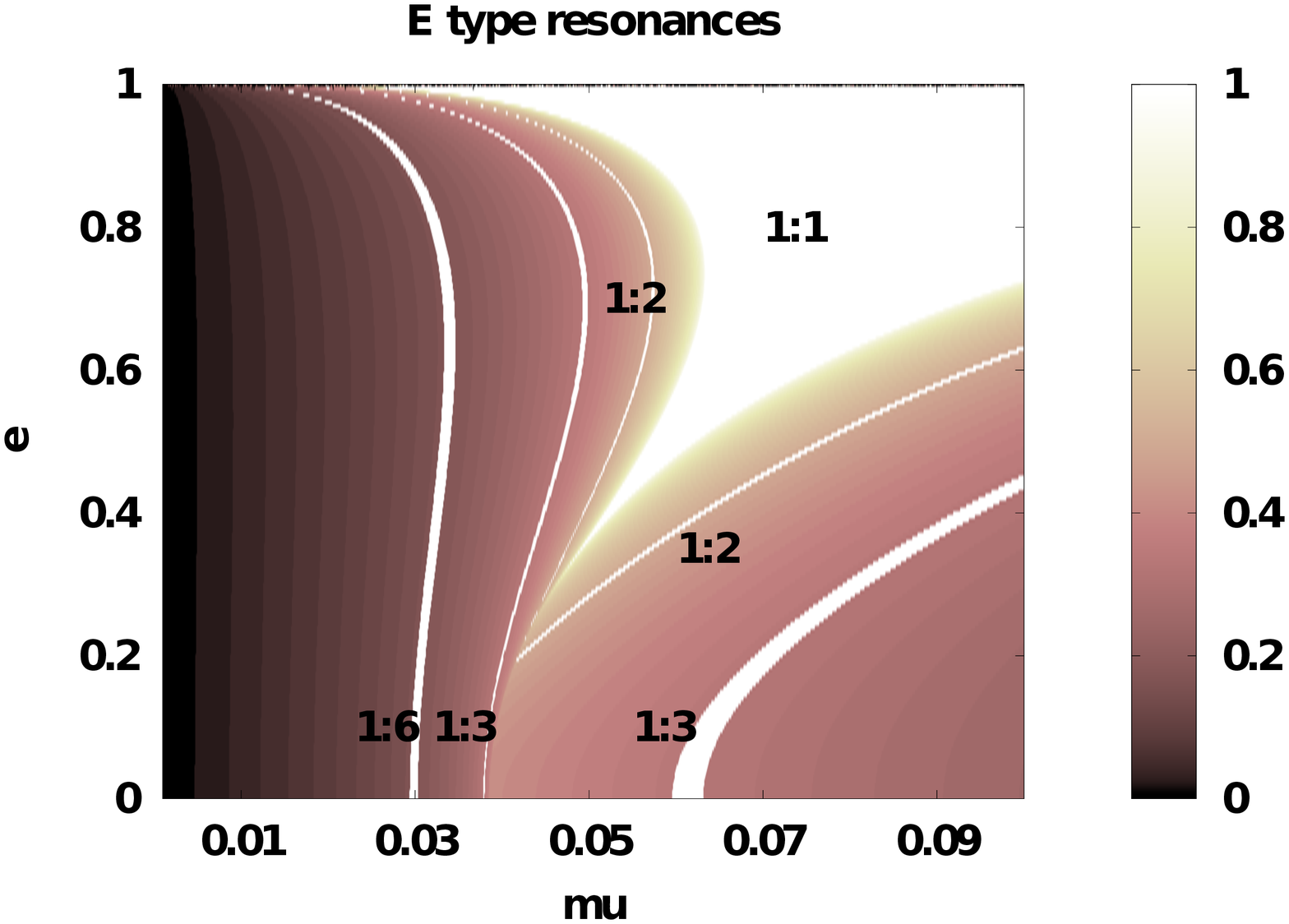}

	\includegraphics[width=0.47\textwidth]{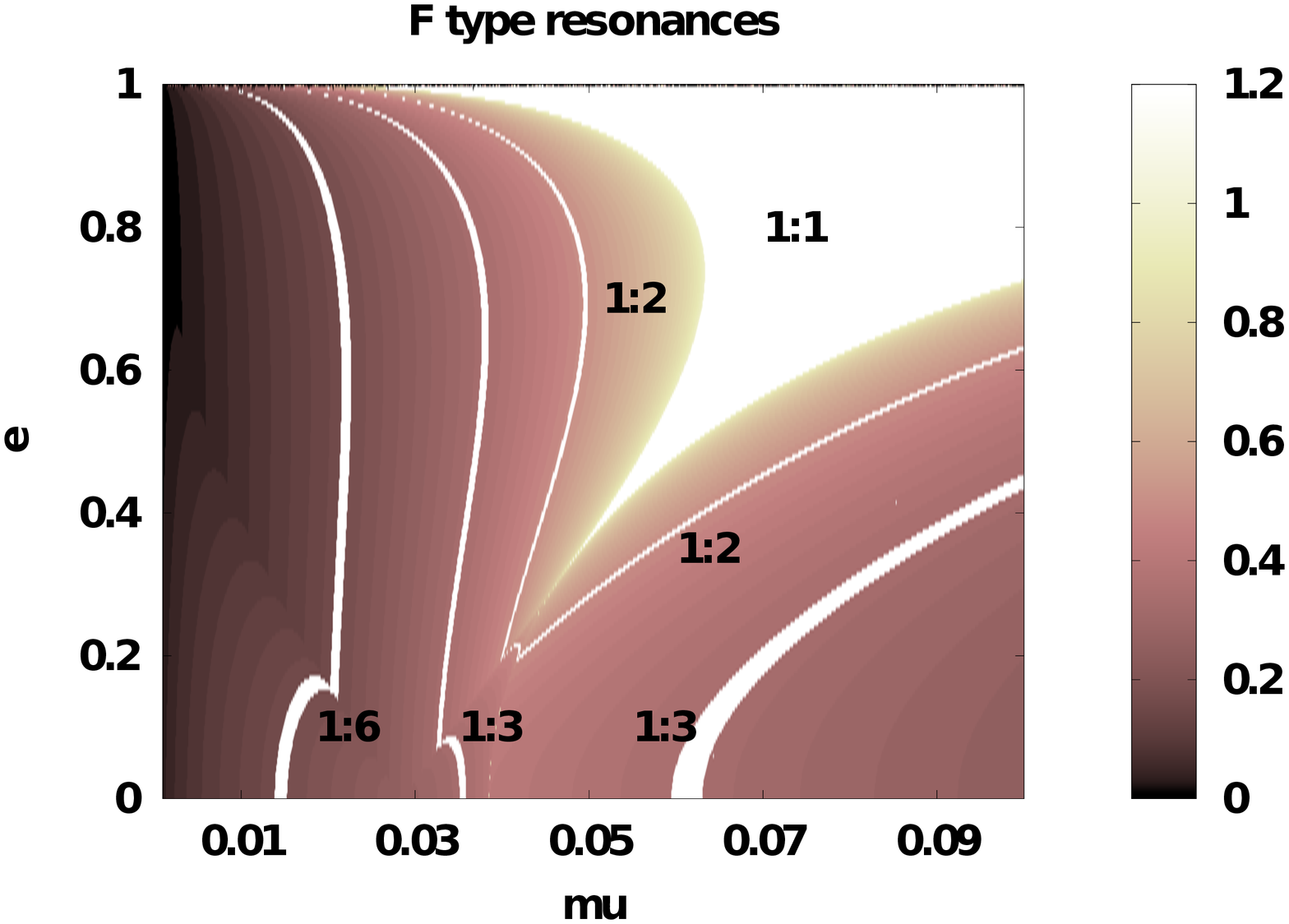}

\caption{Maps for the D, E, and F types of resonances. For further details see the caption of Fig. \ref{ABC}.}

\label{DEF}

\end{figure}

By computing the frequencies via Floquet's theory, we could map the resonances on the $\mu,e$ plane. The results are shown in Figs \ref{ABC} and \ref{DEF}. These extends the results of previous investigations \citep{b2,b4}. For a better visualization, we mapped the reciprocal of the frequency ratios given in Table \ref{Resonances}. Thus the A type 1:2 resonance in the top panel of Fig. \ref{ABC} corresponds to the A type 2:1 resonance in \citet{b2}. Fig. \ref{ABC} shows the frequency ratios of the types A, B, and C, and Fig. \ref{DEF} the same for the types D, E, and F, with some highlighted resonance curves.

\begin{table}

\begin{center}

\caption{Types of resonances}

\label{Resonances}

\begin{tabular}{@{}c r c l}

\hline

\textbf{A} & $(1-n_l)$&:&$n_l$\\

\textbf{B} & $n_s$&:&$n_l$\\

\textbf{C} & $(1-n_l)$&:&$(1-n_s)$\\

\textbf{D} & $n_s$&:&$(1-n_l)$\\

\textbf{E} & $n_s$&:&$(1-n_s)$\\

\textbf{F} & $n_l$&:&$(1-n_s)$\\

\hline

\end{tabular}

\end{center}

\end{table}

\par

The extended white domain in each panel indicates the positions of 1:1 resonances, where the frequencies become equal. This occurs not only along narrow curves as expected before, but in very extended regions which cover large parts of the $\mu,e$ plane, and follow the boundaries shown in Fig. \ref{CharRoots}. The 1:1 resonance of type A (hereafter A11) corresponds to that region in the $\mu,e$ plane, where at least 2 real characteristic roots appear with negative sign (U1 and U3 regions in Fig. \ref{CharRoots}). The negative real characteristic roots result in frequencies, through the characteristic exponents, equal to $0.5$, and these frequencies are the $n_l$, and $1-n_l$ fequencies, corresponding to the A type resonance.  This can also be seen in Fig. \ref{ImagExp}. 

\par

The 1:1 resonance of types B and C (B11 and C11) covers the unified domain of U2 and U3 of Fig. \ref{CharRoots}, where there are 4 complex characteristic exponents. The 1:1 resonances appear due to the fact that the imaginary parts of the characteristic exponents are equal in pairs. In those regions of the $\mu,e$ plane, where 4 negative real characteristic roots are possible (U1, U2, and U3 in Fig. \ref{CharRoots}), the imaginary parts of the characteristic exponents are all equal to $-0.5$. Thus in these regions we expect A11, B11, C11, D11, E11, and F11 resonances, and the maps of Figs \ref{ABC} and \ref{DEF} actually show this. One can see that $1:1$ resonances cover the whole unstable domain of the $\mu,e$ parameter plane, suggesting that they can be responsible in a large measure for the loss of stability of the point $L_4$ in the ERTBP.

\par

It can be seen from Fig. \ref{DEF} that for those values of $\mu$ and $e$

which correspond to the U2 domain, the D, E, and F resonances have the same character. The reason for this is that in the U2 domain there are also B11 and C11 resonances, meaning that $n_s = n_l$, and $1-n_l=1-n_s$, and it follows (see Table \ref{Resonances}) that D=E=F. Moreover, these resonances are also equal to 1/A (compare the A 3:1 and D, E, F 1:3 resonances in Figs \ref{ABC} and \ref{DEF}).

\par

The resonance curves of the A, B, and D types in the S1 and S2 stability domains

are in good agreement with the results obtained in \citet{b2} by using Rabe's equation \citep{b8}, and the A and E type resonances agree with those determined in \citet{b5} by using the energy-rate method (the E type resonance was computed in both the stable and unstable domains). We note that the C and F type resonance curves show peculiar shapes in the stable domain. This is in connection with the phenomenon that resonance curves do not behave regulary on the boundary of the stability region. Crossing the boundary they break, making difficult to find a good fitting function. The most striking examples can be seen in the cases of the C and F type resonances in Figs \ref{ABC} and \ref{DEF}.

\begin{figure}

	\includegraphics[width=0.35\textwidth,angle=270]{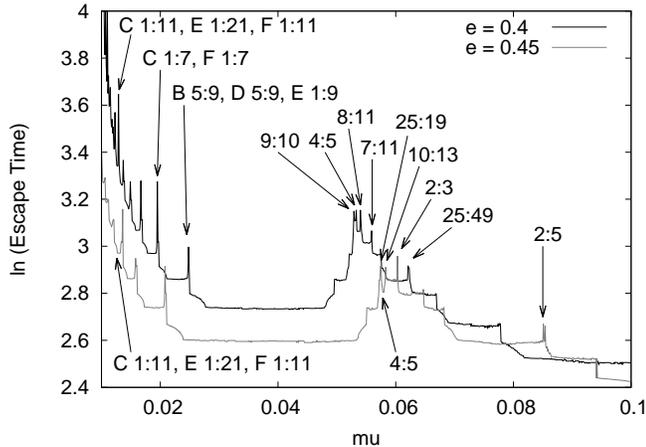}

\caption{Resonances corresponding to longer escape times of the test particle in the unstable domains for $e=0.4$ and $0.45$. Resonances without the letter of the type refer to A, D,  E, and F multiple resonances. For example, 4:5 means A 5:4, D 4:5, E 4:5, and F 4:5 resonances.}

\label{ResonancePeaks}

\end{figure}

We know from stability investigations of the Solar system that resonances can protect, but can also destroy systems of celestial bodies. \citet{b5} suggested that resonances can be responsible for longer escape times of the test particle in the unstable domains of the $\mu,e$ plane. We checked if the peaks in the escape time of the test particle (see Fig. \ref{ResonancePeaks}) could be in connection with some resonances.

\par

Fig. \ref{ResonancePeaks} shows the changing of the escape time of the test particle depending on $\mu$ for $e=0.4$ and $0.45$. Several peaks of the escape time curves can be identified with resonances. Among these protective resonances there are low order (2:3, 2:5, 4:5), and high order (25:49, 1:11, 1:21) resonances as well. There are also many multiple resonances, corresponding to different types at the same time (like 9:10, 7:11 referring to A, D, E, and F types). 

\par

An interesting behaviour is that by changing the eccentricity, protective resonances can become destroying ones, decreasing the escape time of the test particle. For example, for $e=0.4$ the C 1:11, E 1:21, F 1:21, and 4:5 resonances are protective, but for $e=0.45$ they are destroying to which there correspond pits in the escape time curve in Fig. \ref{ResonancePeaks}.


\section[Summary]{Summary}

\label{Summary}

By using Floquet's theory \citep{b9} and following the methods described by \citet{b3} and \citet{b1}, we computed the characteristic roots and  characteristic exponents of infinitesimal motion around the Lagrangian point $L_4$ in the elliptic restricted three-body problem for the domains of the mass parameter $0<\mu \leq 0.5$ and eccentricity $0 \leq e<1$. We numerically integrated the first variational equations of motion around $L_4$ to build up the fundamental matrix, from its eigenvalues we determined the characteristic roots and exponents, and from the latters the frequencies of the periodic components of motions depending on $\mu$ and $e$. According to the properties of the characteristic roots, we distinguished stable and unstable domains of $L_4$ in the $\mu,e$ plane in agreement with former results \citep{b3,b1} . 

\par

We computed the frequencies not only in the stable domains, but also in the unstable regions. We also determined frequencies by the method of fast Fourier transform. The results obtained by the two methods are in good agreement, but the method based on Floquet's theory were applicable in such regions of the $\mu,e$ plane, where FFT did not work (due to short data sets as a consequence of fast escape of the test particle).  

\par

We studied how the librational frequencies in the stable domain depend on the mass parameter and the eccentricity. Based on the frequency - mass parameter relationship of the circular restrictd three-body problem, we determined fitting functions for the frequencies in the ERTBP which give the frequencies depending on $\mu$ and $e$. These are given in the Appendix.

\par

Between the 4 frequencies of motion around $L_4$ in the ERTBP, there can be 6 types of resonances \citep{b2,b4}. We studied the frequency ratios in the whole investigated $\mu,e$ plane. We found that 1:1 resonances occur not only along narrow curves, but in very extended regions, and different types of 1:1 resonances cover the whole unstable region in the $\mu,e$ plane. Different types of resonances correspond to different types of characteristic roots and characteristic exponents in the $\mu,e$ plane. 

\par

The escape time of the test particle in the unstable domain, depending on $\mu$ and $e$, can also be related to resonances. There are protective resonances, where the escape times are longer at the corresponding values of $\mu$  and $e$. These resonances can be responsible for the longer escape times of the test particle from $L_4$. However, by changing the parameters ($e$ for a given $\mu$) the character of the resonance may change resulting in shorter escape time of the test particle.


\section*{Acknowledgments}


\appendix

\section[Fitting the frequencies]{Fitting the frequencies}

The $n_l$ curve is separated to two sides, $n_{l,l}$ and $n_{l,r}$,  and two different parameter sets were used for their fit. Between the two sides, $n_l$ is constant, $n_l = 0.5$. The $\mu_l$ and $\mu_r$ coordinates of the breakpoints of the $n_l$ curve for given values of $e$ can be computed  from the polynomial functions  given in \citet{b4}: 

\[
  \mu_l=0.0285955-0.0555801e+0.0090947e^2+0.0341118e^3
\]

\[
\qquad \qquad -0.0163862e^4,
\]

\[
\mu_r=0.0285955+0.0577951e+0.0026645e^2+0.0234761e^3
\]

\[
\qquad \qquad -0.0750853e^4.
\]

The fitting functions for the frequencies are (changing the notations of Section \ref{SectionFit} for convenience, and keeping only the fitted coefficients)

\begin{equation}
n_s(\mu) = \sqrt{|A_1 +A_2\sqrt{27(\mu+A_3)^2 -5.75} | }
\nonumber
\end{equation}

\begin{equation}
n_l(\mu) = \left\{
\begin{array}{l}
 \sqrt{|B_1 +B_2\sqrt{27(\mu+B_3)^2 -5.75} | }\\
\qquad \rmn{if} \  \mu < \mu_l \\ [5pt]
  0.5 \quad \rmn{if} \  \mu_l < \mu < \mu_r  \\ [5pt]
 0.5\sqrt{|C_1 +C_2\sqrt{27(\mu+C_3)^2 -5.75} | } +C_4 \\
\qquad \rmn{if} \  \mu > \mu_r \\
\end{array}
\right.
\nonumber
\end{equation}

where the $A_i$, $B_i$, $C_i$ coefficients are polynomials of the eccentricity 

\[
A_i(e) = \sum_j a_{i,j} \cdot e^j, \ B_i(e) = \sum_j b_{i,j} \cdot e^j, \ C_i(e) = \sum_j c_{i,j} \cdot e^j,
\] 

except $B_1$, where we used an exponential fitting function: $B_1 = b_{1,1}\exp{(b_{1,2}\,e)} + b_{1,3}$.

\subsection{Fitting parameters for $\mathbf{n_s}$}

\begin{equation}
n_s(\mu) = \sqrt{|A_1 +A_2\sqrt{27(\mu+A_3)^2 -5.75} | }
\nonumber
\end{equation}

\begin{equation}
\begin{array}{r c l}
	A_1 & = &a_{1,1}+a_{1,2}e+a_{1,3}e^2+a_{1,4}e^3+a_{1,5}e^4+a_{1,6}e^5+\\
		& & +a_{1,7}e^6+a_{1,8}e^7+a_{1,9}e^8\\
	A_2 & = &a_{2,1}+a_{2,2}e+a_{2,3}e^2+a_{2,4}e^3+a_{2,5}e^4+a_{2,6}e^5+\\
		& & +a_{2,7}e^6+a_{2,8}e^7+a_{2,9}e^8\\
	A_3 & = &a_{3,1}+a_{3,2}e+a_{3,3}e^2+a_{3,4}e^3+a_{3,5}e^4+a_{3,6}e^5+\\
		& & +a_{3,7}e^6+a_{3,8}e^7\\
\end{array}
\end{equation}

\begin{equation}
\begin{array}{r c l}
	a_{1,1} &=& -0.501002169796915\\
	a_{1,2} &=& -0.00465201800792375\\
	a_{1,3} &=& 1.39021245993088\\
	a_{1,4} &=& -7.65704222733915\\
	a_{1,5} &=& 58.9694647229867\\
	a_{1,6} &=& -184.999091047886\\
	a_{1,7} &=& 268.30016373139\\
	a_{1,8} &=& -184.282471189398\\
	a_{1,9} &=& 48.6925607416981\\
 & &\\
	a_{2,1} &=& -0.500999575602288\\
	a_{2,2} &=& 0.0037393921033813\\
	a_{2,3} &=& -0.883321850087518\\
	a_{2,4} &=& 7.55148105681823\\
	a_{2,5} &=& -59.917802755943\\
	a_{2,6} &=& 200.631458043772\\
	a_{2,7} &=& -317.115523217342\\
	a_{2,8} &=& 240.11269595431\\
	a_{2,9} &=& -70.8037648777467\\
& &\\
	a_{3,1} &=& -0.499999914619729\\
	a_{3,2} &=& -0.000238236951561991\\
	a_{3,3} &=& -0.0864876445889037\\
	a_{3,4} &=& 0.190520373359826\\
	a_{3,5} &=& -1.35033175168857\\
	a_{3,6} &=& 3.62324711720567\\
	a_{3,7} &=& -3.8958192249963\\
	a_{3,8} &=& 1.52170604628458
\end{array}
\nonumber
\end{equation}

\subsection{Fitting parameters for $\mathbf{n_{l,l}}$}

\begin{equation}
n_{l,l}(\mu) = \sqrt{|B_1 +B_2\sqrt{27(\mu+B_3)^2 -5.75} | }
\nonumber
\end{equation}

\begin{equation}
\begin{array}{r c l}
	B_1 & = &b_{1,1}\exp{(b_{1,2}e})+b_{1,3}\\
	B_2 & = &b_{2,1}+b_{2,2}e+b_{2,3}e^2+b_{2,4}e^3+b_{2,5}e^4+b_{2,6}e^5+\\
		& & + b_{2,7}e^6+b_{2,8}e^7\\
	B_3 & = &b_{3,1}+b_{3,2}e+b_{3,3}e^2+b_{3,4}e^3+b_{3,5}e^4\\
\end{array}
\end{equation}

\begin{equation}
\begin{array}{r c l}
	b_{1,1} &=& -0.283143803539808\\
	b_{1,2} &=& -8.80063684576695\\
	b_{1,3} &=& -0.217255471020594\\	
 & &\\
	b_{2,1} &=& 0.50036132088182\\
	b_{2,2} &=& -1.50595975540227\\
	b_{2,3} &=& 10.115844050892\\
	b_{2,4} &=& -63.0931155920938\\
	b_{2,5} &=& 285.264900151338\\
	b_{2,6} &=& -666.727927110675\\
	b_{2,7} &=& 752.16230243293\\
	b_{2,8} &=& -322.241229133452\\
	\\
& &\\
	b_{3,1} &=& -0.499930735128903\\
	b_{3,2} &=& 0.15970524654665\\
	b_{3,3} &=& -0.398529593818837\\
	b_{3,4} &=& 0.561419591971936\\
	b_{3,5} &=& -0.303312424339199
\end{array}
\nonumber
\end{equation}

\subsection{Fitting parameters for $\mathbf{n_{l,r}}$}

\begin{equation}
n_{l,r}(\mu) = 0.5\sqrt{|C_1 +C_2\sqrt{27(\mu+C_3)^2 -5.75} | } + C_4
\nonumber
\end{equation}

\begin{equation}
\begin{array}{r c l}
	C_1 & = &c_{1,1}+c_{1,2}e+c_{1,3}e^2+c_{1,4}e^3+c_{1,5}e^4+c_{1,6}e^5\\
	C_2 & = &c_{2,1}+c_{2,2}e+c_{2,3}e^2+c_{2,4}e^3\\
	C_3 & = &c_{3,1}+c_{3,3}e^2\\
	C_4 & = &c_{4,1}+c_{4,2}e+c_{4,3}e^2+c_{4,4}e^3\\
\end{array}
\end{equation}

\begin{equation}
\begin{array}{r c l}
	c_{1,1} &=& -0.769762964976643\\
	c_{1,2} &=& 10.5612231908263\\
	c_{1,3} &=& -54.5231687808546\\
	c_{1,4} &=& -10.8373555491553\\
	c_{1,5} &=& 892.817398534855\\
	c_{1,6} &=& -1921.57814251696\\
 & &\\
	c_{2,1} &=& 1.13820612112604 \\
	c_{2,2} &=& -11.2718231877969\\
	c_{2,3} &=& 64.2577426648982\\
	c_{2,4} &=& -113.455649510537\\
& &\\
	c_{3,1} &=& -0.499867875893531\\
	c_{3,3} &=& -0.0859479289026476\\	
& &\\
	c_{4,1} &=& 0.252209443579731\\
	c_{4,2} &=& 3.53516545799185\\
	c_{4,3} &=& -17.6970689361691\\
	c_{4,4} &=& 30.0317028979599\\
\end{array}
\end{equation}

\bsp

\label{lastpage}

\end{document}